\documentclass[final]{article}
\usepackage{graphicx}
\usepackage{rotating}
\usepackage{amssymb,amsmath}
\usepackage{wasysym}
\usepackage{mathptmx}
\usepackage[numbers]{natbib}
\usepackage{subscript}
\usepackage{gensymb} 
\usepackage{fancyhdr}
\usepackage{lastpage}
\usepackage{abstract}
\usepackage{hyphenat}
\usepackage{marvosym}
\usepackage{ragged2e}
\usepackage{subscript}
\usepackage{booktabs}
\usepackage{tabularx}
\usepackage{xcolor}
\usepackage[version=3]{mhchem}

\usepackage[vcentering,dvips,left=3.2cm,right=3.2cm,top=4cm,bottom=4cm]{geometry}

\usepackage{authblk}

\definecolor{AtomictAngerine}{rgb}{1.0, 0.6, 0.4}
\definecolor{UniBlue}{RGB}{83,121,170}

\usepackage{hyperref}
\hypersetup{
  breaklinks,
  pdfstartview=XYZ,
  bookmarks=true,
  colorlinks=true,
  linkcolor=UniBlue,
  urlcolor=UniBlue,
  citecolor=UniBlue,
  pdftex,
  bookmarks=true,
  linktocpage=true,   
  hyperindex=true
}

\usepackage{fancyhdr}
\renewenvironment{abstract}{%
  \hfill\par
    {\noindent\bfseries Abstract\par}
    \noindent\rule{\textwidth}{0pt}}
    {\par\noindent\rule{\textwidth}{0pt}
}

\makeatletter

\renewcommand\@maketitle{%
  \hfill
  \hspace*{-2em}
  \begin{minipage}{\textwidth}
  \vskip 2em
  \let\footnote\thanks 
  {\noindent\LARGE\bfseries\@title\par}
  \vskip 1.5em
  {\noindent\large\@author\par}
  \end{minipage}
  \vskip 1em \par
}
\makeatother

\newenvironment{acknowledgements}
    {\par\addvspace{17pt}\small\rmfamily\noindent {\bfseries Acknowledgements}%
    \small\rmfamily\noindent}%
    {}%

\newcommand{\note}{\footnote}
\newcommand{\affiliation}{\affil}

\makeatletter

\setcitestyle{square}

\begin{document}

\title{\nohyphens{Development of Thermal Kinetic Inductance Detectors suitable for X-ray spectroscopy}}

\author[1,2]{A.~Giachero\note{Corresponding author \Telefon~+39 02-6448-2456, \Letter~Andrea.Giachero@mib.infn.it}}
\author[3]{A.~Cruciani}
\author[4]{A.~D'Addabbo}
\author[5]{P.~K.~Day}
\author[6,7]{S.~Di~Domizio}
\author[1,2]{M.~Faverzani}
\author[2]{E.~Ferri}
\author[8,9]{B.~Margesin}
\author[10,3]{M.~Martinez}
\author[11,9]{R.~Mezzena}
\author[10,3]{L.~Minutolo}
\author[1,2]{A.~Nucciotti}
\author[1,2]{A.~Puiu}
\author[3]{M.~Vignati}

\affiliation[1]{Universit\`{a} di Milano-Bicocca, Milano, Italy}
\affiliation[2]{INFN - Sezione di Milano-Bicocca, Milano, Italy}
\affiliation[3]{INFN - Sezione di Roma1, Roma, Italy}
\affiliation[4]{INFN - Laboratori Nazionali del Gran Sasso (LNGS), Assergi (AQ), Italy}
\affiliation[5]{Jet Propulsion Laboratory, Pasadena, CA, U.S.A.}
\affiliation[6]{Dipartimento di Fisica - Universit\`{a} degli Studi di Genova, Genova, Italy}
\affiliation[7]{INFN - Sezione di Genova, Genova, Italy}
\affiliation[8]{Fondazione Bruno Kessler (FBK), Trento, Italy}
\affiliation[9]{INFN - Trento Institute for Fundamental Physics and Applications (TIFPA), Trento, Italy}
\affiliation[10]{Dipartimento di Fisica - Sapienza Universit\`{a} di Roma, Roma, Italy}
\affiliation[11]{Dipartimento di Fisica, Univerist\`{a} di Trento, Povo, Trento, Italy}


\maketitle

\justify\begin{abstract}
We report on the development of Thermal Kinetic Inductance Detectors (TKIDs) suitable to perform X-ray spectroscopy measurements. The aim is to implement MKIDs sensors working in thermal quasi-equilibrium mode to detect X-ray photons as pure calorimeters. The thermal mode is a variation on the MKID classical way of operation that has generated interest in recent years. TKIDs can offer the MKIDs inherent multiplexibility in the frequency domain, a high spatial resolution comparable with CCDs, and an energy resolution theoretically limited only by thermodynamic fluctuations across the thermal weak links. 

Microresonators are built in Ti/TiN multilayer technology with the inductive part thermally coupled with a metal absorber on a suspended SiN membrane, to avoid escape of phonons from the film to the substrate. The mid-term goal is to optimize the single pixel design in term of superconducting critical temperatures, internal quality factors, kinetic inductance and spectral energy resolution. The final goal is to realize a demonstrator array for a next generation thousand pixels X-ray spectrometer.  

In this contribution, the status of the project after one year of developments is reported, with detailed reference to the microresonators design and simulations and to the fabrication process.
\end{abstract}

\section{Introduction}
The thermal mode is a variation on the MKID classical way of operation that has generated interest in recent years~\cite{TArgonne,TSB}. The responsivity of a MKID is related to the derivative of the complex conductivity $\sigma=\sigma_1-j\sigma_2$ with respect to the quasiparticle density $n_{qp}$~\cite{MattisBardeen}. In non-equilibrium mode the excess quasiparticles $d\sigma/d n_{qp}$ is due to an external pair breaking source. In 2008 J. Gao \textit{et al.}~\cite{Thermal2008} demonstrated the equivalence between \textit{athermal} production of quasiparticles and pair-breaking due to thermal effect. A temperature change can produce an identical increase of quasiparticle population of an external pair-breaking. This way of operation is called \textit{thermal quasi-equilibrium mode}. This equivalence is routinely used to statically measure the response and uniformity of large arrays of resonators as a function of temperature. In simple terms the effect of a small variation in temperature leads to a change in the surface impedance due to a change in kinetic inductance $L_{kin}$. 

\begin{figure}[!t] 
 \begin{center}
    \includegraphics[clip=true,width=\textwidth]{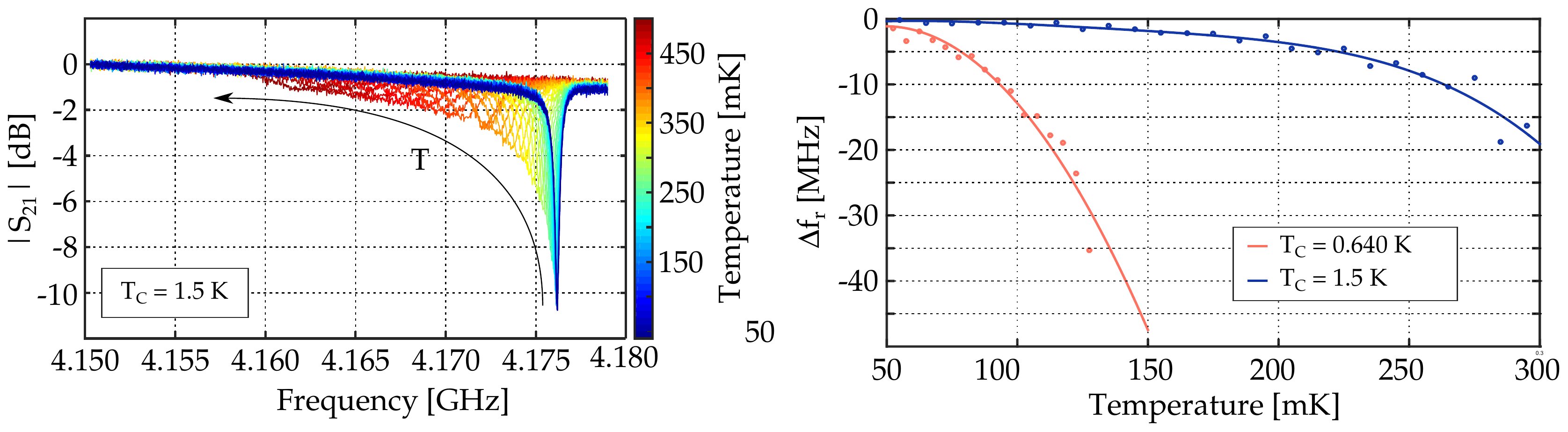}
  \end{center}
 \caption{\label{fig:thermal} (left) Resonance shift characterization as a function of the temperature. (right) Decrease of resonance frequency $f_r$ with increasing temperature $T$. The variation $\Delta f_r$ is steeper with lower critical temperatures. Plots edited from A. Giachero {\it et al.}~\cite{MiB2}}
\end{figure}

The KIDS\_RD project plans to exploit this equivalence to use absorber-coupled superconducting resonators to detect X-ray photons in a thermal quasi-equilibrium manner (pure calorimeter). In this way the amplitude and phase shifts of the microwave transmitted signal depends on the increase in equilibrium thermal quasiparticle population due to the bath temperature variation (figure \ref{fig:thermal}, left).  The absorber material must have a very high stopping power, needed in order to avoid loss of energy and to keep the thermal capacity as low as possible. Elements with high atomic number (i.e. Gold or Bismuth) are preferable. The resonator and absorber could not be in direct electrical contact. A membrane of SiN\textsubscript{x} must be used to thermally isolate the absorber from substrate and from the resonator itself, avoiding energy losses. The energy flow between the resonator and absorber is mediated by the phonons through the membrane.

This operation mode can be very sensitive by using devices with critical temperature low enough, to maximize the temperature dependance (figure \ref{fig:thermal}, right), and operating at very low temperature to minimize the absorber thermal capacity. The energy resolution is theoretically limited only by thermodynamic heat fluctuations across the thermal weak links: $\Delta E_{FWHM} \simeq 2.355\,\sqrt{kT^2 C}$, where $k$ is the Boltzmann constant, $C$ the heat capacity of the absorber and $T$ the bath temperature. Considering the intrinsic properties of a metal absorber (i.e. Gold) $200~\mu\mbox{m}\times 200~\mu\mbox{m}\times 2~\mu\mbox{m}$ working at $T = 50$~mK, it is possible to achieve resolutions around 1~eV. 

The best resolution achieved with TKID~\cite{TSB} is $\Delta E = 75$\,eV at 6\,keV. In this case microresonators were fabricated from a 200\,nm thick substoichiometric TiN\textsubscript{x} film, with 500\,nm thick superconducting Ta absorber suspended on a free-standing Si\textsubscript{3}N\textsubscript{4} membrane. This resolution was limited by saturation effect due to a not optimized coupling between the thermal bath and the detector. A better thermal design could improve this value, moreover by finding the optimal tradeoff between the frequency response and the critical temperature is possible to have a better response as a function of the temperature change. In this work the selected design solutions and fabrication process to obtain thermal KIDs able to fulfil these requirements are presented.

\section{Microresonators Design}
\begin{figure}[!t] 
 \begin{center}
    \includegraphics[clip=true,width=\textwidth]{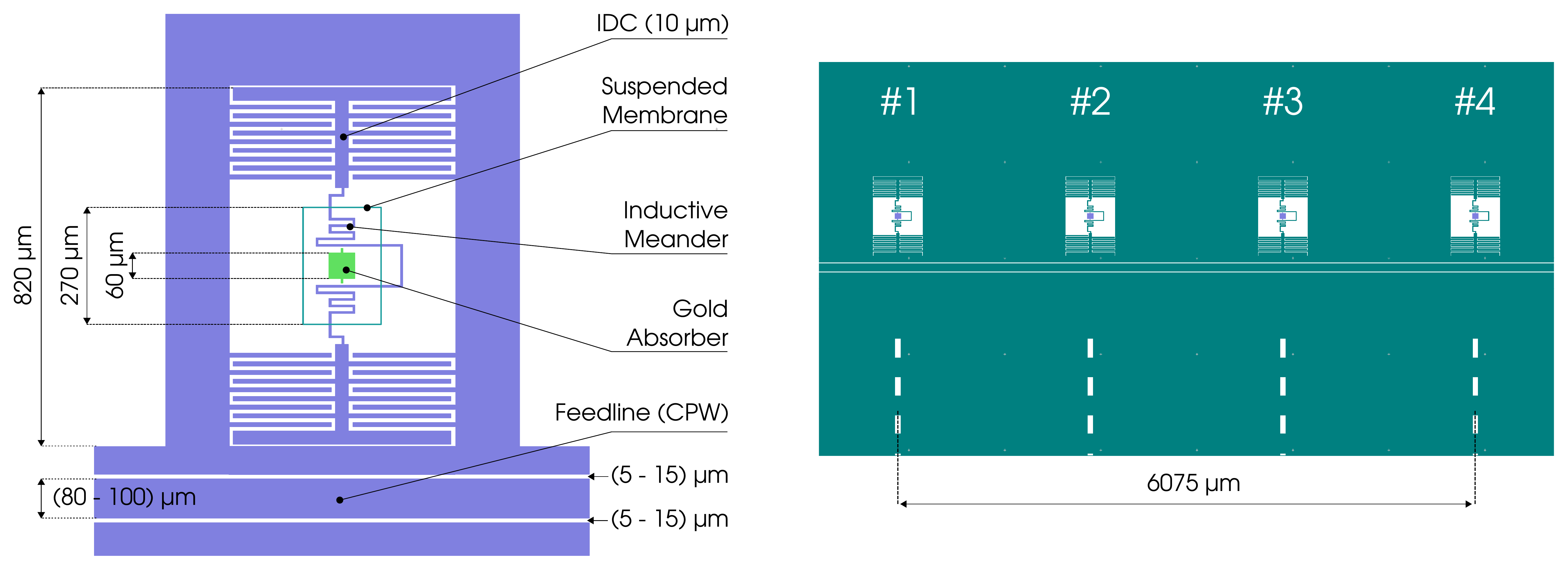}
  \end{center}
 \caption{\label{fig:res} (left) Example of resonator geometry with gold absorber thermally coupled to the inductive meander and both suspended on a SiN membrane. (right) Chip array accommodating 4 microresonators.}
\end{figure}

The developed resonator geometry is in the lumped element form, evolution of the one presented in a previous work~\cite{MiB0}. It consists of two interdigital capacitors (IDC) connected with a coplanar waveguide (CPW) that works as inductor. The resonator is capacitively coupled on one side to a coplanar waveguide (CPW) line that is used for the readout (figure \ref{fig:res}, left). The spacing and width of the conductors of the IDCs is designed to be around 10\,$\mu$m. This spacing is intended to reduce two-level systems (TLS) noise associated with amorphous dielectric layers at surfaces by decreasing the surface layer to volume ratio of the capacitors. The strength of the coupling, which sets the coupling quality factor $Q_c$  is determined by the width of the gap between the CPW line and the resonator. Different width for the CPWs and for the gaps have been designed in order to have different resonator configurations, combination of different kinetic impedances ($L_k = 12,\,20~\mbox{and}~30\,\mbox{pH/sq}$, corresponding to three different critical temperatures) and nominal quality factors ($Q = 5\cdot 10^3 , 15\cdot 10^3~\mbox{and}~40\cdot 10^3$).  

The TKID is realized by thermally coupling a gold absorber to the microresonator inductive meander. A gold thickness of 2\,$\mu$m provides a stopping power close to 99.99998\% (99.927\%) full stopping for 2\,keV electrons (photons). Three different variants have been designed with different geometries and different distance absorber-inductor. The first is $2 \times 60 \times 60\,\mu\mbox{m}^3$ absorber 15\,$\mu$m close to the meander, the second is $2 \times 30 \times 90\,\mu\mbox{m}^3$ absorber 30\,$\mu$m close, while the last one is similar to the first with the addiction of $2\times 10 \times 5\,\mu\mbox{m}^3$ fingers on the sides faced to the meander, with a minimum distance of 5\,$\mu$m from it. The purpose of these fingers is to increase the thermal coupling between the absorber and the inductor. These different configurations will be used to study possible influences on the microresonator performances caused by the absorber. The absorber and part of the meander are suspended on a silicon nitrate membrane 1\,$\mu$m thick to minimize phonon exchange and to provide a finite thermal conductance to the bath. The resulting free-standing area is $270 \times 180\,\mu\mbox{m}^2$ for all the configurations. 

\begin{figure}[!t] 
 \begin{center}
   \includegraphics[clip=true,width=0.85\textwidth]{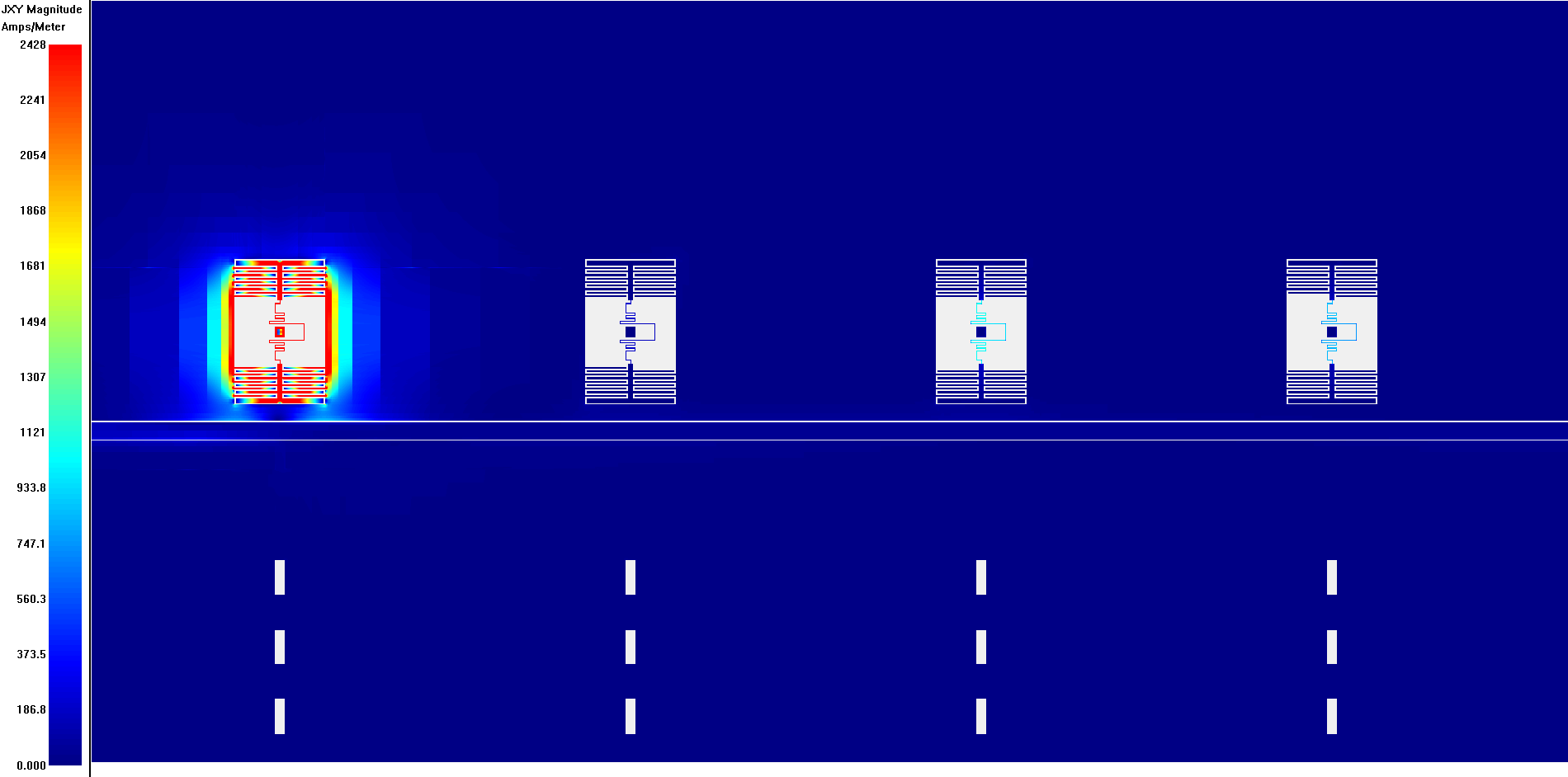}
  \end{center}
 \caption{\label{fig:simulation} A simulation showing the time-averaged magnitude of the current flowing through the array when the signal frequency matches the resonant frequency of one microresonator (the first on the left in figure).}
\end{figure}

Each array accommodates $4 \times 1$ thermal KIDs, with resonant frequencies in the 4-6.5\,GHz range, for a total chip size of $19800 \times 7800\,\mu\mbox{m}^2$ (figure \ref{fig:res}, right). The designed prototypes cover all combinations of kinetic inductances, quality factors and absorbers described above, for a total of 69 microresonator arrays produced on the same wafer. All the designed solutions have been simulated and studied by using a 3D planar RF electromagnetic analysis software (Sonnet). The simulations take into account both the kinetic inductance of the superconductor as well as the resistance of the gold absorber. In addition to the estimation of the usual parameters (e.g. coupling quality factor, resonant frequency, line impedance), this analysis was useful to estimate the degradation of internal quality factor due to the proximity of the normal metal and to determine the best distance between meander and absorber. An example of an entire array with 4 microresonator simulated at the same time is shown in figure \ref{fig:simulation}.

To obtain three different values of kinetic inductance (i.e. three different critical temperatures $T_c$) three wafers with three different superconducting film are currently in production for a total of 207 microresonator arrays. After a deep characterization the best performing resonator configuration will be implemented in a larger array.

\section{Fabrication Process}
The main important requirement for our superconducting films is a very low transition temperature $T_c$ tunable in the range within 500\,mK and 1.5\,K. Sub-stoichiometric titanium nitride (TiN\textsubscript{x}) is a material that has been popular in the KID community for some time due to its high resistivity and tunable $T_c$. However, since $T_c$ is a very strong function of the nitrogen content and a precise control of the nitrogen incorporation process is difficult to maintain, the control of the targeted $T_c$ from run-to-run and across a wafer area is difficult to achieve~\cite{NIST}. To overcome these limitations we consider a different approach by employing a multilayer film made of pure Ti and stoichiometric TiN~\cite{MiB1}. In fact, exploiting the proximity effect the $T_c$  of a superconducting material can be reduced by the superimposition of a normal metal or a metal with a super-conductive transition at lower temperature. As shown in previous works a good reproducibility and uniformity in the films production can be obtained~\cite{MiB1,MiB2}. 

\begin{figure}[!t] 
 \begin{center}
    \includegraphics[clip=true,width=\textwidth]{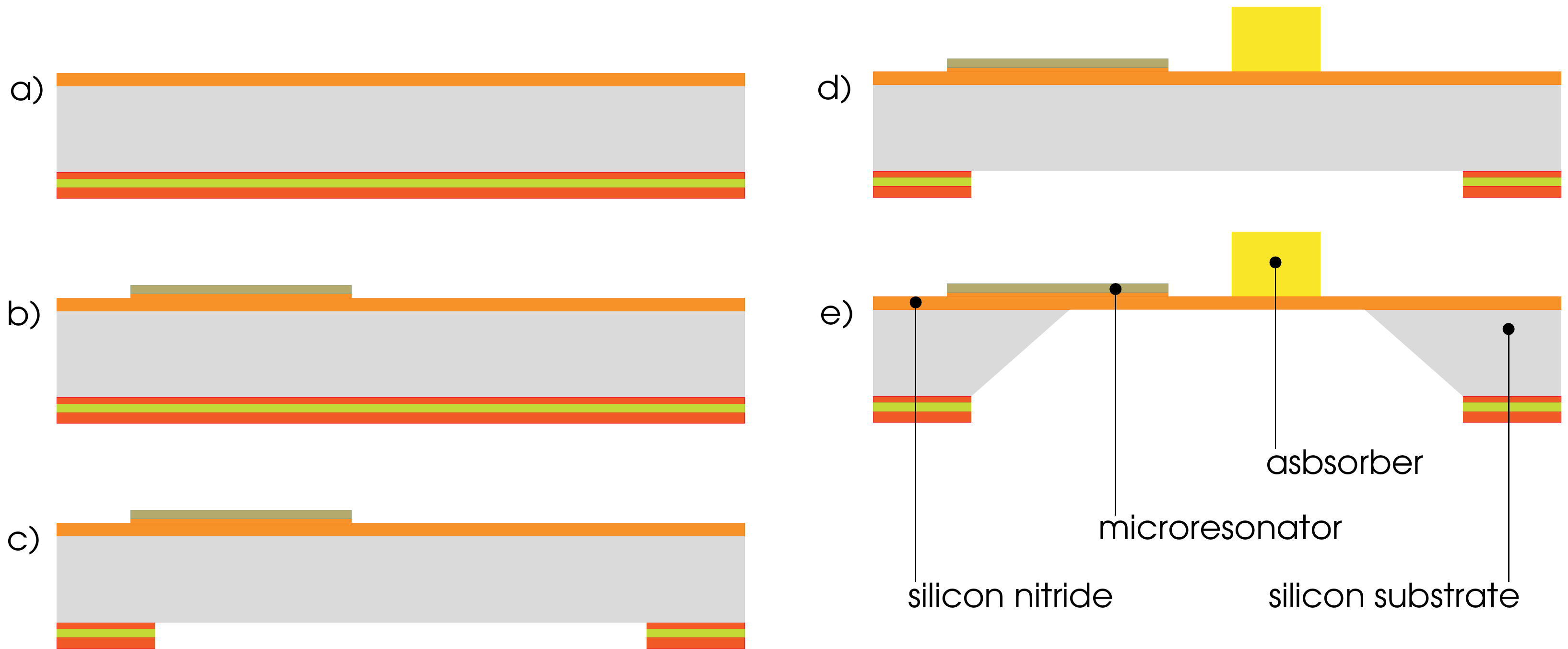}
  \end{center}
 \caption{\label{fig:production} Outline of the micromachined fabrication process carried out at the MicroSystems Technology (MST) Research Unit at the FBK.}
\end{figure}

The detectors are fabricated on a 6'' double side polished 625\,$\mu$m thick 100 oriented high resistive 5000\,$\Omega$cm p-type silicon wafers. The fabrication steps are (figure \ref{fig:production}): a) deposition of the silicon nitrate membrane (SiN). This step consists on deposition of a composite hard mask consisting in 300\,nm of thermal oxide, 150\,nm of stoichiometric silicon nitride followed by 300\,nm of a medium temperature oxide obtained by vapour deposition of tetramethyl orthosilicate (TEOS) on the wafer backside. On the wafer front side a 725 nm thick low stress silicon nitride is deposited by Plasma Enhanced CVD; b) The multilayers film is then deposited as superimposition of a number of bilayers Ti/TiN by sputtering for a total thickness of around 100\,nm at 350\,$^{\circ}$C in which the microresonators are defined and etched. Following the developed recipe~\cite{MiB1}, the critical temperature can be tuned by keeping fixed the Ti thickness and by properly varying the TiN thickness. Specific details are reported in table \ref{tab:thickness}; c) In a second lithography step the etch window for the bulk micromachining is defend and opened in the hard mask on the wafer backside; d) A thin titanium Ti/Au seed layer is deposited on the front side and with a 10\,$\mu$m thick photoresist a mask is defined for the galvanic deposition of the 2\,$\mu$m thick gold absorber; e) At the end the free-standing membrane is defined by removing the silicon under the silicon nitride by bulk silicon etching in a tetra methyl ammonium hydroxide: water solution (TMAH). The designed TKIDs are currently in fabrication at the MicroSystems Technology (MST) Research Unit at the Fondazione Bruno Kesler (FBK, Trento, Italy)

\begin{table}[!t]
 \begin{center}
    \begin{tabular}{cccccc}
      \toprule
      Ti & TiN & N layers  & Total & Target $T_C$  & Target $L_s$ \\
      {[nm]}& {[nm]}    &   & thickness  & [K]  & [pH/\Square]\\
      \midrule
      10 & 12 & 5 & 110 & 1.5 & 12\\
      10 & 10 & 5 & 100 & 0.8 & 20\\
      10 & 7 & 6 & 102 & 0.6 & 30\\
      \bottomrule
    \end{tabular}
  \end{center}
  \caption{\label{tab:thickness} Titanium and titanium nitrade thickness used to reach the target critical temperature and sheet inductance.}
\end{table}

\section{Status and future plans}
KIDS\_RD is a project funded by the Italian Institute of Nuclear Physics (INFN) with the aim to develop Thermal Kinetic Inductance Detectors suitable to perform X-ray spectroscopy measurements. The first phase of the project is focused on the single pixel optimization. Several configurations with with different geometry designs have been developed and the detectors are currently in fabrication at the MST micromachining facility at the FBK in Trento. Starting from the end of 2017 and continuing for the entire 2018, a deep detector characterization will be performed. At the end the best performing resonator configuration will be selected and implemented in a larger array. Scalable and resolving X-ray detectors could open plenty of opportunities for ground-breaking and science-enabling measurements in a broad range of disciplines like particle physic, astrophysics, materials science, biological and life sciences, medicine, chemistry, earth science, archeoemetry and cultural heritage.

\begin{acknowledgements}
  This work is carried out in the frame of the KIDS\_RD R\&D project funded by the Istituto Nazionale di Fisica Nucleare (INFN) in the Commissione Scientifica Nazionale 5 (CSN5). We also acknowledge previous support from Fondazione Cariplo, through the project \textit{Development of Microresonator Detectors for Neutrino Physics} (grant \textit{International Recruitment Call 2010}, ref. 2010-2351), and from INFN through the MARE project.

\end{acknowledgements}

\pagebreak

\end{document}